\documentclass[twocolumn]{article}
\usepackage{graphicx}
\usepackage{psfrag}

\newcommand{\myscalebox}[1]{\scalebox{0.4}[0.4]{#1}}

\newcommand{\mysection}[1]{\section{\em #1}}
\begin{document}

\title{Ground-state clusters of two-, three- and four-dimensional $\pm J$
  Ising spin glasses}


\author{Alexander K. Hartmann,\\
{\small  hartmann@theorie.physik.uni-goettingen.de}\\
{\small Institut f\"ur theoretische Physik, Bunsenstr. 9}\\
{\small 37073 G\"ottingen, Germany}\\
{\small Tel. +49-551-395995, Fax. +49-551-399631}}

\date{\today}
\maketitle
\begin{abstract}

A huge number of independent true ground-state configurations 
is calculated for two-, three- and four-dimensional $\pm J$ spin-glass models.
Using the genetic cluster-exact approximation method,  system sizes up to
$N=20^2,8^3,6^4$ spins are treated. 
A ``ballistic-search'' algorithm
 is applied which allows even for large system sizes
to identify clusters  of
ground states which are connected by chains of zero-energy flips of spins. 
The number of clusters $n_C$ diverges with $N$ going to infinity. 
 For all dimensions
 considered here, an exponential increase of $n_C$ appears to be
 more likely than a growth with a power of $N$.
The number of different ground states is found to grow clearly 
exponentially with 
$N$. A zero-temperature
entropy per spin of $s_0=0.078(5)k_B$ (2d), $s_0=0.051(3)k_B$ (3d)
respectively $s_0=0.027(5)k_B$ (4d) is obtained.
 
{\bf Keywords (PACS-codes)}: Spin glasses and other random models (75.10.Nr), 
Numerical simulation studies (75.40.Mg),
General mathematical systems (02.10.Jf). 
\end{abstract}

\mysection{Introduction}

Spin-glass models
\cite{binder86} with discrete distributions of the interactions
are believed to exhibit a rich ground-state ($T=0$) landscape.
So far only for very small systems of $N=4^3$ spins the complete
landscape has been analyzed \cite{klotz94}. In this work a new 
``ballistic search'' algorithm is presented, which allows the
treatment of much larger systems.
As an application,  two-, three-, and four-dimensional 
Edwards-Anderson (EA) $\pm J$
spin glasses  are  investigated. They consist of $N$ spins 
$\sigma_i = \pm 1$, described by the Hamiltonian
\begin{equation}
H \equiv - \sum_{\langle i,j\rangle} J_{ij} \sigma_i \sigma_j \quad .
\end{equation}

The sum runs over all pairs of nearest neighbors.
The spins are placed on  $d=2,3,4$-dimensional  
simple (square/ cubic/ hypercubic)
lattices of linear size $L$ with periodic boundary conditions in
all directions.
Systems with quenched disorder of the interactions (bonds)
are considered. Their possible values are $J_{ij}=\pm 1$ with equal
probability. To reduce the fluctuations, a constraint is imposed, such that 
$\sum_{\langle i,j\rangle} J_{ij}=0$.
Since the Hamiltonian exhibits no external
field, reversing all spins of a {\em configuration} (also called {\em
  state}) $\{\sigma_i\}$  results in a state with the same energy, 
called the {\em inverse} of
$\{\sigma_i\}$. In the following, a spin configuration
 and its inverse are regarded as one single state.

Since the ground-state problem belongs to the class of
NP-hard tasks \cite{barahona82}, 
only algorithms with exponentially increasing running time
are available. Currently it
is possible to obtain a finite number of true ground states per realization
up to $L=56$ (2d), $L=14$ (3d) or $L=8$ (4d) using a special
optimization algorithm \cite{alex-sg234d}.  
For the $\pm J$ model the number of existing ground states $n_{GS}$
per realization, called the
{\em ground-state degeneracy}, grows exponentially
with $N$. The reason is, that there are
usually {\em free} spins, i.e. spins which can be flipped without changing the
energy of the system. A state with $f$ independent free 
spins allows at
least for $2^f$ different configurations of the same energy. 
Currently, it seems to be 
impossible to obtain all ground states for system sizes larger than $L=5$.
To overcome this problem in this work all {\em clusters} 
 of ground states are calculated. A cluster is defined in the
 following way: Two ground state configurations are called {\em
   neighbors} if they differ only by the orientation of one free 
spin. All ground
 states which are accessible through this neighbor relation are
 defined to be in the same cluster. This means, one can travel
 through the ground states of one cluster by flipping only free spins.
With the method presented here, the {\em ballistic search} (BS),
it is not only possible to analyze large ground-state clusters,
furthermore it allows to obtain the
cluster landscape when having only a small subset of all ground states
available. 
Additionally one can  estimate the size of the clusters from this small
number of sample states, as shown later on. 

The number of clusters as a function of system size is also of interest on
its own: for the infinitely-ranged Sherrington-Kirkpatrik
(SK) Ising spin glass a complex configuration-space structure was found
using the replica-symmetry-breaking mean-field (MF) scheme by Parisi
\cite{parisi2}. If the MF scheme is valid for finite-dimensional spin
glasses as well, then the number of ground-state clusters must diverge
with increasing system size. On the other hand the droplet-scaling picture
 \cite{mcmillan,bray,fisher1-2,bovier}
predicts that basically one ground-state cluster (and its inverse) dominates
the spin-glass behavior.
To address this issue  a cluster-analysis was performed
for small systems of one size $L=4$ in three dimensions
\cite{klotz94}.  But an analysis of the size
 dependence of the number of clusters or even an investigation of
 two-/four-dimensional spin glasses has not been carried out before.

By the way, with the method presented here 
it is possible to calculate the
entropy 
$S_0 \equiv \langle \ln n_{GS} \rangle k_B$ even for systems exhibiting a
huge $T=0$ degeneracy.
The symbol $\langle \ldots \rangle$ denotes 
the average over different realizations of the bonds.
Since the number of free spins is extensive,
 $s_0\equiv S_0/N>0$ holds for the $\pm J$ spin glass. 
For three-dimensional spin glasses,
in \cite{morgenstern80} the ground-state entropy was estimated by
computing exact free energies for systems of size $4\times 4\times M$ 
($4\le M \le 10$). In
\cite{kirkpatrick77} a Monte-Carlo simulation and in
\cite{berg93,berg94} multicanonical simulations were used to calculate 
$s_0$. Results for the ground-state entropy of two dimensional system
were obtained with similar methods: numerically exact calculations of
finite systems \cite{morgenstern80b,vannimenus77}, Monte-Carlo
simulations\cite{kirkpatrick77,wang88} and analytics methods
\cite{cheung83,blackman91,blackman98}. For $d=4$ the author is not
aware of results for the ground-state entropy.

The paper is organized as follows: First the procedures
used in this work are presented. Then the results for the number of clusters
and the number of ground states as function of $N$ in two, three and
four dimensions are
shown. The last section a summarizes the results.

\mysection{Algorithms}

At first the optimization method applied here is stated. Then, for
illustrating the problem, a simple method for constructing clusters of
ground states is explained. In the main part 
the BS method for identifying clusters in
systems exhibiting a huge degeneracy is presented and it is explained
how this technique can be applied to estimate the size of these clusters.

The basic method used here for the calculation of spin-glass ground
states is the cluster-exact approximation (CEA) algorithm 
\cite{alex2}, which is a
discrete optimization method designed especially for spin glasses. In
combination with a genetic algorithm \cite{pal96,michal92} this method is
able to calculate true ground states \cite{alex-sg234d}. 
Using this technique one does not
encounter ergodicity problems or critical
slowing down like in algorithms which are based 
on Monte-Carlo methods. Genetic CEA
was already utilized to examine the ground states of
two-, three- and four dimensional 
 $\pm J$ spin glasses by calculating a small
number of 
ground states per realization \cite{alex-sg234d}, while in this work
the emphasize is on the study of the cluster landscape. Therefore,
many ground states per random
sample have to be obtained. 
Since the algorithm calculates only one independent ground
state per run, a much larger computation effort was necessary.

Once many ground states are calculated the straight-forward method to
obtain the cluster landscape works the following way: 
The construction starts
with one arbitrary ground state. All  its neighbors are added to the
cluster. These neighbors are treated recursively in the same way: All
their neighbors which are yet not included in the cluster are
added. After the construction of one cluster is completed the
construction of the next one starts with a ground state, which has not
been visited so far.

The construction of the clusters needs only linear computer-time as
function of $n_{SG}$ ($O(n_{SG})$), similar to the Hoshen-Kopelman
technique \cite{hoshen76}, because each ground state is visited only
once. Unfortunately the detection of all neighbors, which has to be 
performed at the beginning, is of $O(n_{SG}^2)$
since all pairs of states have to be compared. Even worse, all
existing ground states must have been calculated before. As e.g. a $5^3$
system may exhibit already more than $10^5$ ground states, this
algorithm is not suitable.

The basic idea of the ballistic-search algorithm is to use a {\em test}, which
tells whether two ground states are in the same cluster. The test works
as follows: Given two independent replicas $\{\sigma_i^{\alpha}\}$
and $\{\sigma_i^{\beta}\}$ let $D$ be the set of spins, which are
different in both states: $D\equiv \{i|\sigma_i^{\alpha}\neq
\sigma_i^{\beta}\}$. Now BS tries to build a path of successive
flips of free spins, which leads from  $\{\sigma_i^{\alpha}\}$ to
$\{\sigma_i^{\beta}\}$ while using only spins from $D$.
In the simplest version iteratively a free spin is selected randomly
from $D$, flipped
and removed from $D$. This test does not guarantee to find a path
between two ground states which belong to the same cluster, since it
may depend on the order the spins are selected whether a path is
found or not. It only
finds a path with a certain probability which depends on the size of
$D$. It turns out that the probability decreases monotonically with $|D|$. 
For example for $N=8^3$ the method finds a path in 90\% of all
cases if the two states differ by 34 spins. More analysis can
be found in \cite{alex-ballistic}. 

The algorithm for the identification of clusters using BS works as
follows: the basic idea is to let a ground state represent that part
of a cluster which can be found using BS with a high probability 
by starting at this ground state. If a
cluster is large it has to be represented by a collection of states, 
such that the whole
cluster is ``covered''. For example a typical cluster of a $8^3$ spin
glass consisting of $10^{16}$ ground states is usually represented by
only some few ground states (e.g. two or three). 
A detailed analysis of how many representing
ground states are needed as a function of cluster and system size
can be found in \cite{alex-ballistic}. The
algorithm holds in memory a set of clusters consisting each of a set of
representing configurations. At the beginning the cluster set is empty.
Iteratively all available ground states $\{\sigma_i\}$
are treated: For all representing configurations the BS algorithm 
tries to find a path to the current ground state or to its inverse. 
If no path is found, a new cluster is
created, which is represented by the actual configuration treated. If
$\{\sigma_i\}$ is found to be in exactly one cluster nothing special
happens. If $\{\sigma_i\}$ is found to be in more than one cluster all
these clusters are merged into one single cluster, which is now represented by
the union of the states which have represented all clusters affected
by the merge.

The BS identification algorithm has some advantages in comparison
with the straight-forward method:
since each ground-state configuration represents many ground
states, the method does not need to compare all pairs
of states. Each state is compared only to a few number of representing
configurations. Thus, the computer time needed for the calculation 
grows only a little
bit faster than $O(n_{SG}n_C)$ \cite{alex-ballistic}, 
where $n_C$ is the number of clusters which is much smaller than
$n_{SG}$.  Consequently, large sets of ground states, 
which appear already for small system sizes like
$L=5$, can be treated. Furthermore, the
ground-state cluster landscape of even larger
systems can be analyzed, since it is sufficient to calculate a small
number of ground states per cluster. One has to ensure that
really all clusters are found, which is simply done by calculating 
enough states, but this  is still only a tiny fraction of all ground
states  \cite{alex-ballistic}. Also one has to be sure that all
clusters are identified correctly. This is not guaranteed immediately,
since for two ground states belonging to the same cluster there is
just a certain probability that a path of free flipping spins
connecting them is found. But this poses no problem, because once at least
one state of a cluster has been found, many more states can be obtained
easily by just performing a $T=0$ Monte-Carlo simulation starting with
the initial state. By
increasing the number of states available more and more, the
probability that all clusters have been identified correctly
approaches very quickly one. Detailed tests can be found in
\cite{alex-ballistic}. For all results presented here, the number of
available ground states has been increased so far, such that each cluster
has be identified correctly with a probability of more than 0.99.

Once all ground states are grouped into clusters, 
their sizes have to be obtained to calculate the total number of
states and
the entropy. If only some ground states per cluster are available, the
size cannot be evaluated by simply counting the states.
Then a variant of BS is used to perform this task. Given a
state $\{\sigma_i\}$, 
free spins are flipped iteratively, but each spin not more than
once. During the iteration additional free spins may be generated or destroyed.
When there are no more free spins left, the process stops.
One counts the number of spins that has been flipped. By averaging
over several tries and several ground states of a cluster one obtains an
average value, denoted with  $l_{\max}$. It can be shown that this quantity
represents the size $n_C$ of a cluster very well and is more accurate
than simpler measures such as
the average number of static free spins. By analyzing 
all ground states of small systems 
a $n_C=2^{\alpha l_{\max}}$  behavior is found, with 
$\alpha\in[0.85,0.93]$ depending on the dimension of the system. These
results will be exposed in the next section.
 A similar method for estimating the cluster
sizes is presented in \cite{hed2000}. There
 three heuristic fitting-parameters are needed, but they are universal
 for all system dimensions. 

\mysection{Results}

First, the results for three-dimensional systems are given. In the
second and third part two- and four-dimensional spin glasses are
investigated. 

In 3d,  for system sizes $L=3,4,5,6,8$ large numbers of independent ground
states were calculated using genetic CEA. Usually 1000
different realizations of the disorder were considered. Tab. I shows
the number of realizations $n_R$ and the
number of independent runs $r$ per realization for the different
system sizes $L$. For the small systems sizes (and for 100
realizations of $L=5$) many runs plus an additional local search
were performed to calculate {\em all}
ground states. For the larger sizes $L=5,6,8$ the number of
ground states is too large, so it is only possible to try to
calculate at least one ground state per cluster. It 
 is highly probable that all clusters
were detected, except for $L=8$, where for about 25\% of the realizations
some small cluster may have been missed \cite{alex-ballistic}. 
This problem is not related to the design of the
ballistics search method. 
It is due to the enormous computational effort needed for generating
the ground states of the 
largest systems, so only a restricted number of runs can be performed.
 Since the probability that a certain cluster  is found in a run
of the genetic CEA algorithm decreases with the size of the
cluster \cite{alex-cea-analysis}, ground states belonging to
small clusters occur only rarely. 
Even by doubling the number of runs for
$L=8$, this fraction is estimated to fall only to 20 \%.

The
ground states were grouped into clusters using the ballistic-search
algorithm. The number of states per cluster was sufficiently large, so
that only with a probability of less than $10^{-3}$ some configurations
from a large cluster may be
mistaken for belonging to  different clusters \cite{alex-ballistic}.
The average number $n_C$ of clusters is shown in the fourth column
of Tab I. In Fig. 1 the result is
shown as a function of the number  $N$ of spins. 
By visualizing the results using a double-logarithmic plot (see
inset) one realizes
that $n_C$ seems to grow faster than any power of $N$. The larger
slope in the linear-logarithmic plot 
for small systems may be a finite-size effect. Additionally, for
$L=8$ there is a large probability that some small clusters are missed,
explaining the smaller slope there. 
Summarizing, our data favor an exponential increase of $n_C(N)$. 

To calculate the ground-state entropy, the size of the clusters have
to be known. For the small systems, this can be done  just by
counting. For larger system sizes it is not possible to obtain all
states, so the method using the dynamical number $l_{\max}$ 
of free spins is
applied, as explained before. In Fig. \ref{figEstSize} the cluster
size for small systems is shown as a function of $l_{\max}$ with
a logarithmically scaled y-axis. A $n_C=2^{\alpha l_{\max}}$ 
dependence is visible very well, yielding $a=0.90(5)$.

By summing up all cluster
sizes for each realization the ground-state degeneracy $n_{GS}$ is obtained.
Its average is shown in the fifth column of the table.
The quantity is plotted in Fig. \ref{figNumStates}
as a function of $N$. The exponential growth is obvious. 

The result for the average ground-state entropy per spin is 
shown in the last column of Tab. I. The number for $L=4$ 
is within two standard deviations of 
$s_0=0.073(7)k_B$ which was found in \cite{klotz94}, 
where 200 realization were treated. 
By fitting a function of the
form $s_0(L)=s_0(\infty)+a*L^{-\beta}$ a value of 
$s_0(\infty)=0.0505(6)k_B$ is obtained. In \cite{morgenstern80} 
$s_0=0.04(1)k_B$ was estimated for systems with periodic
boundary conditions only in two directions, which may be the reason
for the smaller result. The value found by a Monte-Carlo
simulation $s_0=0.062k_B$ \cite{kirkpatrick77} for systems of size
$20^3$ is much larger. The deviation is presumably caused by the fact
that it was not possible to obtain true ground states for systems of
that size, i.e. too many states were found.
 The results from multicanonical simulations $s_0=0.046(2)k_B$
\cite{berg93} and $s_0=0.0441(5)k_B$ \cite{berg94}
are a little bit lower than the
results obtained here. This may indicate that not all ground states
are found using that simulation procedure.

The result for the entropy does not suffer from the fact, that some
ground-state clusters may have been missed for $L=8$: the
probability for finding a cluster using genetic CEA grows with the size of
the cluster \cite{alex-cea-analysis}. 
This implies that the clusters, which may have been missed, are
considerably small, so the influence on the result is
negligible. The largest source of uncertainty is caused by the
assumption, that the size of a cluster grows like $2^{\alpha
 l_{\max}}$. The error of the constant $\alpha$ enters linearly 
the result of the entropy. 
To estimate the influence of this approximation, for the three smallest
systems sizes, where the entropy was obtained exactly, $s_0$ was
calculated using estimated cluster sizes as well. For all three
cases the result was equal to the exact values within error bars.
The final result quoted here is $s_0=0.051(1)$.

Now we concentrate on two-dimensional systems. 
For system sizes $L=5,7,10,14,20$ large numbers of independent ground
states were calculated using genetic CEA, up to $10^4$ runs per
realization were performed. Usually 1000
different realizations of the disorder were considered, except for $L=20$,
where only 96 realizations could be treated
For the small systems sizes $L=5,7$, many runs plus an additional 
local search were performed to calculate {\em all}
ground states. For the larger sizes $L=10,14,20$ the number of
ground states is too large, so we restrict ourselves to
calculate at least one ground state per cluster. The probability that
some clusters were missed is higher for two dimensions than for the
$d=3$ case, because the ground-state degeneracy grows faster with the
system size: for small systems sizes $L\le 10$
it is again highly probable that all clusters
have been obtained. For $L=14$ some small
clusters may have been missed for about 30\% of all realizations, 
while for $L=20$ this fraction raises even to
60\%. This is due to the enormous computational effort needed for the 
largest systems. 
For the $L=20$ realizations a total computing time of more than 2
CPU-years was consumed on a cluster of Power-PC processors running with
80MHz.

The results for $d=2$ are shown in Tab. II. The number of clusters
$n_C$ as a function of system size is plotted in
Fig. \ref{figNumClustersTwoD}. Again it is more likely that $n_C$
grows exponential than an algebraic growth. 

Similar to the $d=3$ case, the cluster sizes $V$ can be obtained directly
for small systems. For estimating $V$ in larger systems, again the
$\alpha$ parameter has been obtained. The average size of a cluster as
a function of $l_{max}$ is shown in Fig. \ref{figEstSizeTwoD}
resulting in $\alpha=0.85(5)$. With this parameter the ground-state
degeneracy as a function of $N$ can be calculated, see Fig.
\ref{figNumStatesTwoD}. Similar to the $d=3$ case, the exponential
growth is obvious. The resulting entropy is shown in the inset. By a
finite-size extrapolation to the infinite system, a value of
$s_0=0.078(5)$ is obtained.
In \cite{morgenstern80b} 
$s_0\approx 0.075k_B$ was estimated by using a recursive method to
obtain numerically exact free energies up to $L=18$. 
The result of
$s_0\approx 0.07k_B$ found in \cite{vannimenus77} is even slightly lower.
The value found by a Monte-Carlo
simulation $s_0\approx0.1k_B$ \cite{kirkpatrick77} for systems of size
$80^2$ is much larger. The deviation is presumably caused by the fact
that it was not possible to obtain true ground states for systems of
that size, i.e. too many states were visited.
Recent results are more accurate: by applying the replica Monte Carlo  
method \cite{wang88} a value of $s_0=0.071(7)$ was obtained. A  
transfer matrix calculation \cite{cheung83} resulted in  
$s_0=0.0701(5)$. By using a Pfaffian method $s_0=0.0704(2)$  
\cite{blackman91} respectively $s_0=0.0709(4)$ \cite{blackman98}  
was obtained.   
The most recent values are smaller than the entropy found in this  
work. The reason may be that larger systems could be treated (up to  
$L=256$ in \cite{blackman91,blackman98}), while here an extrapolation  
has been performed with systems of size $L\le 20$. At least,  
the value $s_0[L=22]=0.079(1)$ is comparable to the value of  
$s_0[L=32]=0.0780(8)$ found in \cite{blackman91}.   
Additionally, the fact that for the other works  
the number of antiferromagnetic bonds fluctuates from sample to  
sample while it is kept fixed here may have an influence as well.  
This was tested by calculating ground states   
for small systems $(L\le 10)$,   
where each bond has a probability 0.5 of being (anti-) ferromagnetic.  
In this case the entropy turned out to to 5-10 \% below the values  
found above. For large system sizes, which are out of range  
for the method presented here, this effect should decrease.   

In the last part we turn to four-dimensional $\pm J$ spin glasses. 
Because of the huge computational effort, $N=6^4$ is 
the largest size which could be considered and a reasonable statistics
could be only obtained for $L=5$, since one $L=6$ run takes several
CPU weeks. For details, see Tab. III.

The number of clusters as a function of $N$ is displayed in Fig. 
\ref{figNumClustersFourD}. Here, even more clusters seem to have been missed
than in the two-, and three-dimensional cases. But again, the
data-basis is large enough that an exponential increase of the number
of clusters seems possible.

The dependence of the cluster size on the number of flips of free
spins could be studied only for the smallest system size. Even for
$L=4$ the number of ground states can grow beyond $10^6$, preventing a
reliable analysis. From the $L=3$ data (see
Fig. \ref{figEstSizeFourD})
$a=0.93(3)$ has been estimated.

In the final figure (Fig. \ref{figNumStatesFourD}) the resulting
degeneracy is shown. Here, the small numbers of ground states, which
could be calculated with reasonable effort, already have an influence
on the results. For the largest size, the exponential growth of the
number of ground states with system size is not visible. 
 Please note that in general the average $n_{GS}$ is dominated by few
samples having a large number of ground states. For $L=6$, because of
the small number of realizations, these realizations were not
generated within 10 samples. This explains the deviation from the
exponential growth. 

For the entropy (see inset of Fig.\ref{figNumStatesFourD} ), 
rare samples have less influence since the
logarithm of the number of states is averaged. Consequently, the value of
$s_0=0.027(5)$, which again was obtained by a finite-size scaling fit,
 is much more reliable.

As we have seen, 
the $\alpha$-parameter increases with growing dimension. That means
that the spins contributing to the ground-state degeneracy become more
and more independent, the limit $\alpha=1$ corresponds to the case
were all free spins do not interact with each other. This can be
understood from the decrease of the ground-state entropy. From $d=2$
to $d=4$ $s_0$ drops from 0.078 to 0.027. Thus, 
with growing dimension the number of spins contributing
to the ground state degeneracy decreases quickly, so it becomes less
likely that these spins are neighbors. This effect is
stronger than the increase of the number of neighbors per spin 
from 4 in $d=2$ to 8 in $d=4$. 

\mysection{Conclusion}

True ground states of two-, three- and four-dimensional 
$\pm J$  spin glasses have been
calculated using genetic cluster-exact approximation. 
For each realization many independent ground states have been
obtained, leading to an enormous computational effort: several
months of running 32  PowerPC processors on a parallel computer were
necessary. Clusters of ground states have been
 investigated, which are defined to be the sets
of ground-state configurations which can be accessed from each other
by flipping only  free spins. The ballistic-search method has been
presented, which
allows the fast identification of very large clusters. It can be
assured easily that the ground-state clusters found in this way have
been identified correctly.
It should be pointed out that this method is not a tool for the
{\em calculation} of ground states of large systems, but it allows for a
detailed {\em analysis} of highly degenerate ground-state landscapes.
Indeed, it is possible
to calculate clusters of systems when only a small fraction of 
their states is available. The method should be extendable to similar
clustering-problems. A variant of the technique is used to estimate the
size of clusters.

Ground-state clusters for systems of size up to $L=20$ (2d), $L=8$ (3d)
and $L=6$ (4d) have been
calculated. It means that, in the case of three-dimensions,
these realizations are ten times larger and have
$10^{12}$ times more ground states than the systems treated in
\cite{klotz94}. For the other dimensions similar studies even have not
been performed before at all. 
The number of clusters and the degeneracy as a function of
the number of spins $N$ were evaluated.
 It appears that both
quantities are growing exponentially with $N$ for all three cases
$d=2,3,4$. Consequently, it seems 
unlikely that even larger systems  can be treated accordingly 
in the near future. The ground-state
entropy per spin was found to be $s_0=0.078(5)k_B$ (2d),
$s_0=0.051(1)k_B$ (3d) respectively $s_0=0.027(5)k_B$ (4d). It should be
stressed that the result for the entropy does not depend on the
way a cluster is defined. The specific definition given here is only a
tool, which allows the treatment of systems exhibiting a huge
ground-state degeneracy. If ground states had colors, they could be
grouped according their colors as well instead of performing a
clustering  according their neighbor relationship.

With the method presented here, its is only possible to study the
top level clustering of the ground states. It is not
possible to find substructures within the clusters. This kind of enhanced
analysis can be performed with  other methods \cite{hed2000}. Even
when applying these other techniques, the
ballistic search method is still necessary, since  the
cluster landscape has to be obtained in advance.
There, the ballistic-search clustering is applied
 to guarantee that a ground-state landscape is sampled thermodynamically
correct, see also \cite{alex-equi}.

\mysection{Acknowledgements}

The author thanks K. Battacharya, G. Hed, E. Domany and D. Stauffer
 for interesting discussions. 
He is thankful to M. Otto for critical reading the manuscript.
He was supported by the Graduiertenkolleg
``Modellierung und Wissenschaftliches Rechnen in 
Mathematik und Naturwissenschaften'' at the
{\em In\-ter\-diszi\-pli\-n\"a\-res Zentrum f\"ur Wissenschaftliches
  Rechnen} in Heidelberg and the
{\em Paderborn Center for Parallel Computing}
 by the allocation of computer time. The author obtained financial
 support from the DFG ({\em Deutsche Forschungsgemeinschaft}) under
 grant Zi209/6-1.

\clearpage
\onecolumn

\newcommand{\captionNumClusters}
{Number $n_C$ of ground-state clusters as a function of system size
  $N$ for $d=3$. The inset shows the same data using 
  a double-logarithmic scale. Lines are guide to the eyes only.}

\newcommand{\captionEstSize}
{Average size $V$ of a cluster ($d=3$) as a function of average dynamic number
  $l_{\max}$ of
  free spins (see text) for three-dimensional $\pm J$ spin glasses of
 system sizes $L=3,4,5$, where all ground have been obtained. A
 $V=2^{0.9l_{\max}}$ dependence is found, indicated by a line.}

\newcommand{\captionNumStates}
{Number $n_{GS}$ of ground-states ($d=3$) as a function of system size $N$. The
  number of states grows exponentially with the number of spins. Line
  is guide to the  eyes only. The inset displays the ground-state
  entropy per spin as a function of $L$. The line shows a fit
  extrapolating $s_0$ to the infinite system which yields
$s_0(\infty)=0.0505(6)k_B$.}

\newcommand{\captionNumClustersTwoD}
{Number $n_C$ of ground-state clusters for two-dimensional $\pm J$ spin
  glasses as a function of system size
  $N$. The inset shows the same data using 
  a double-logarithmic scale. Lines are guide to the eyes only.}

\newcommand{\captionEstSizeTwoD}
{Average size $V$ of a cluster as a function of average dynamic number
  $l_{\max}$ of
  free spins (see text) for two-dimensional $\pm J$ spin glasses of
 system sizes $L=5,7$, where all ground have been obtained. A
 $V=2^{0.85l_{\max}}$ dependence is found, indicated by a line.}

\newcommand{\captionNumStatesTwoD}
{Number $n_{GS}$ of ground-states for two-dimensional $\pm J$ spin
  glasses as a function of system size $N$ (with $\alpha=0.8$). The
  number of states grows exponentially with the number of spins. Lines
  are guide to the  eyes only. The inset displays the ground-state
  entropy per spin as a function of $L$. The line shows a fit
  extrapolating $s_0$ to the infinite system which yields
$s_0(\infty)=0.078(5)k_B$.}

\newcommand{\captionNumClustersFourD}
{Number $n_C$ of ground-state clusters for four-dimensional $\pm J$ spin
  glasses as a function of system size
  $N$. The inset shows the same data using 
  a double-logarithmic scale. Lines are guide to the eyes only.}

\newcommand{\captionEstSizeFourD}
{Average size $V$ of a cluster as a function of average dynamic number
  $l_{\max}$ of
  free spins (see text) for four-dimensional $\pm J$ spin glasses of
 system sizes $L=3$, where all ground have been obtained. A
 $V=2^{0.93l_{\max}}$ dependence is found, indicated by a line.}

\newcommand{\captionNumStatesFourD}
{Number $n_{GS}$ of ground-states for four-dimensional $\pm J$ spin
  glasses as a function of system size $N$ (with $\alpha=0.93$). The
  number of states grows exponentially with the number of spins. Lines
  are guide to the  eyes only. The inset displays the ground-state
  entropy per spin as a function of $L$. The line shows a fit
  extrapolating $s_0$ to the infinite system which yields
$s_0(\infty)=0.027(5)k_B$.}

\begin{table}
\begin{tabular}{lrrrrr}
L & $n_R$ & $r$ & $C$ & $n_{GS}$ & $s_0/k_B$ \\\hline
3 & 1000 & 1000 & 1.79(3) & $2.6(2) \times 10^1$ & 0.0842(16)\\
4 & 1000 & $10^4$ & 2.58(6) & $2.1(1) \times 10^2$ & 0.0627(09)\\
5 & 100 & $10^5$ & 4.3(3) & $1.3(4) \times 10^4$ & 0.0519(18)\\
5 & 1000 & 3000 & 3.8(1) & $2.1(3) \times 10^4$ & 0.0560(08)\\
6 & 1000 & 5000 & 6.6(3) & $1.3(3) \times 10^7$ & 0.0535(05)\\
8 & 192 & $2\times10^4$ & 24 (2) & $1.8(1.7) \times 10^{16}$ & 0.0520(07) 
\end{tabular}
\caption{For each system size L ($d=3$): number $n_R$ of realizations, 
number $r$ of independent runs per realization, average
  number $C$ of clusters per realization, average ground state
  degeneracy $n_{GS}$ and the average entropy per spin $s_0$.}
\end{table}
\begin{table}

\begin{tabular}{lrrrrr}
L & $n_R$ & $r$ & $C$ & $n_{GS}$ & $s_0/k_B$ \\\hline
5 & 1000 & 1000 & 1.79(3) & $3.2(2) \times 10^1$ & 0.1041(17)\\
7 & 1000 & $10^4$ & 2.58(6) & $5.4(5) \times 10^2$ & 0.0916(13)\\
10 & 1000 & $10^4$ & 4.3(3) & $7.6(4) \times 10^5$ & 0.0868(10)\\
14 & 1000 & 3000 & 3.8(1) & $1.5(9) \times 10^{12}$ & 0.0863(07)\\
20 & 96 & 5000 & 6.6(3) & $5.1(4.9) \times 10^{25}$ & 0.0854(20)\\
\end{tabular}
\caption{For each system size L ($d=2$): number $n_R$ of realizations, 
number $r$ of independent runs per realization, average
  number $C$ of clusters per realization, average ground state
  degeneracy $n_{GS}$ and the average entropy per spin $s_0$.}
\end{table}

\begin{table}
\begin{tabular}{lrrrrr}
L & $n_R$ & $r$ & $C$ & $n_{GS}$ & $s_0/k_B$ \\\hline
3 & 1000 & $5000$ & 2.99(9) & $2.7(2) \times 10^2$ & 0.0510(07)\\
4 & 455 & $5000$ & 5.2(3) & $9(1) \times 10^5$ & 0.0394(07)\\
5 & 457 & $1000$ & 9.9(5) & $7(7) \times 10^{14}$ & 0.0358(03)\\
6 & 10 & 100 & 15(4) & $3(2) \times 10^{20}$ & 0.0319(16)\\
\end{tabular}
\caption{For each system size L ($d=4$): number $n_R$ of realizations, 
number $r$ of independent runs per realization, average
  number $C$ of clusters per realization, average ground state
  degeneracy $n_{GS}$ and the average entropy per spin $s_0$.}
\end{table}
\clearpage
\twocolumn

\begin{figure}[htb]
\begin{center}
\myscalebox{\includegraphics{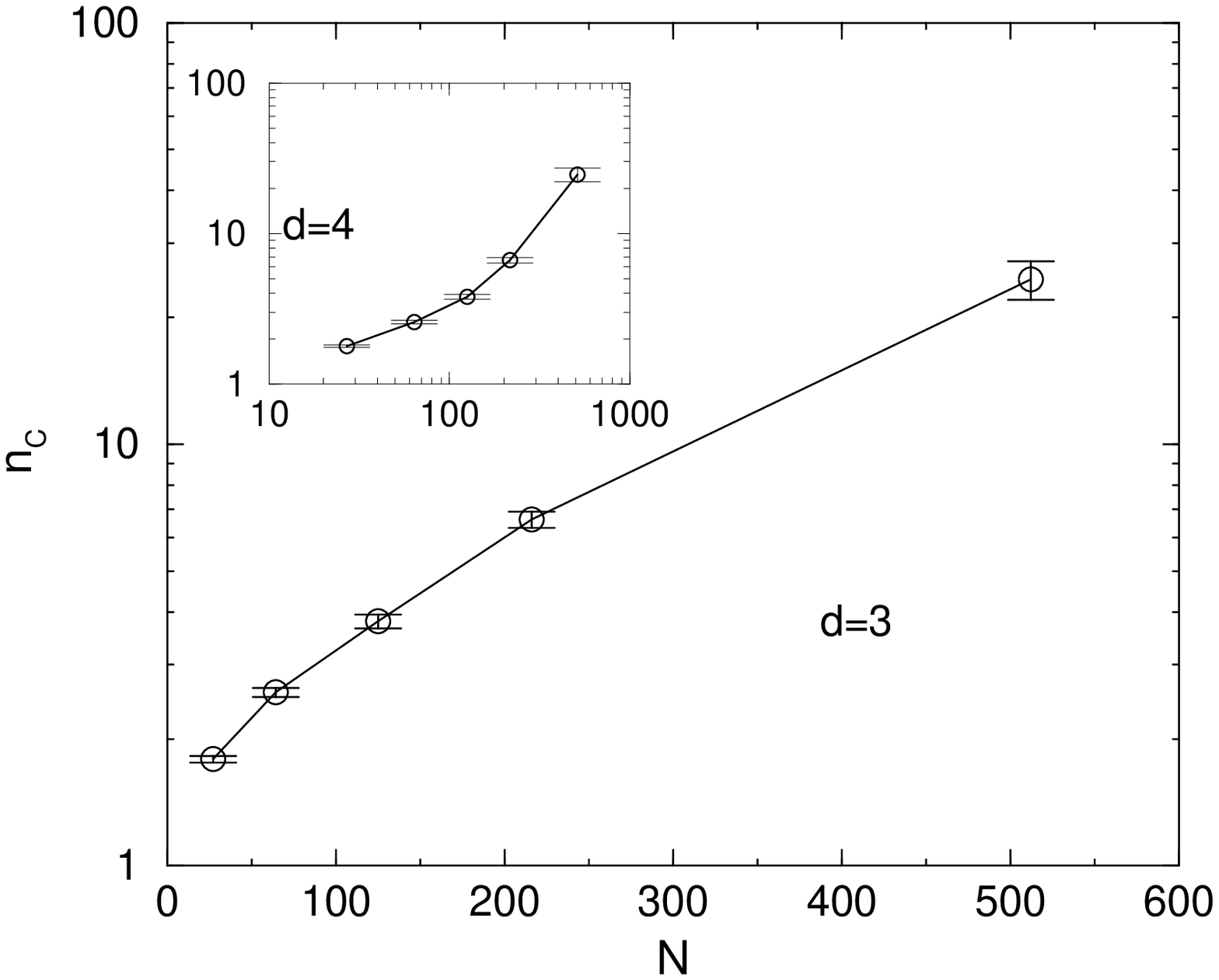}}
\end{center}
\caption{\captionNumClusters}
\label{figNumClusters}
\end{figure}

\begin{figure}[htb]
\begin{center}
\myscalebox{\includegraphics{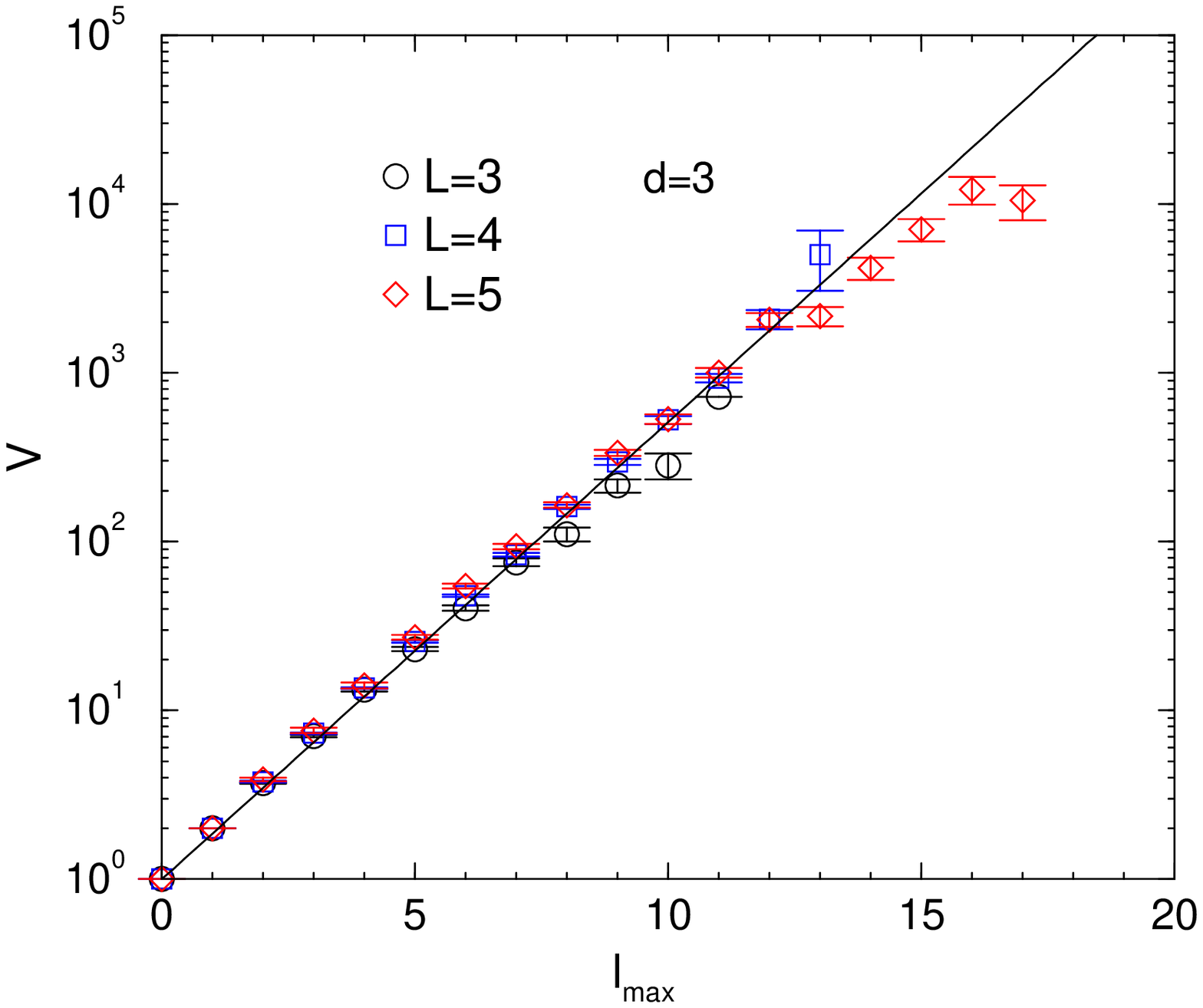}}
\end{center}
\caption{\captionEstSize}
\label{figEstSize}
\end{figure}

\begin{figure}[htb]
\begin{center}
\myscalebox{\includegraphics{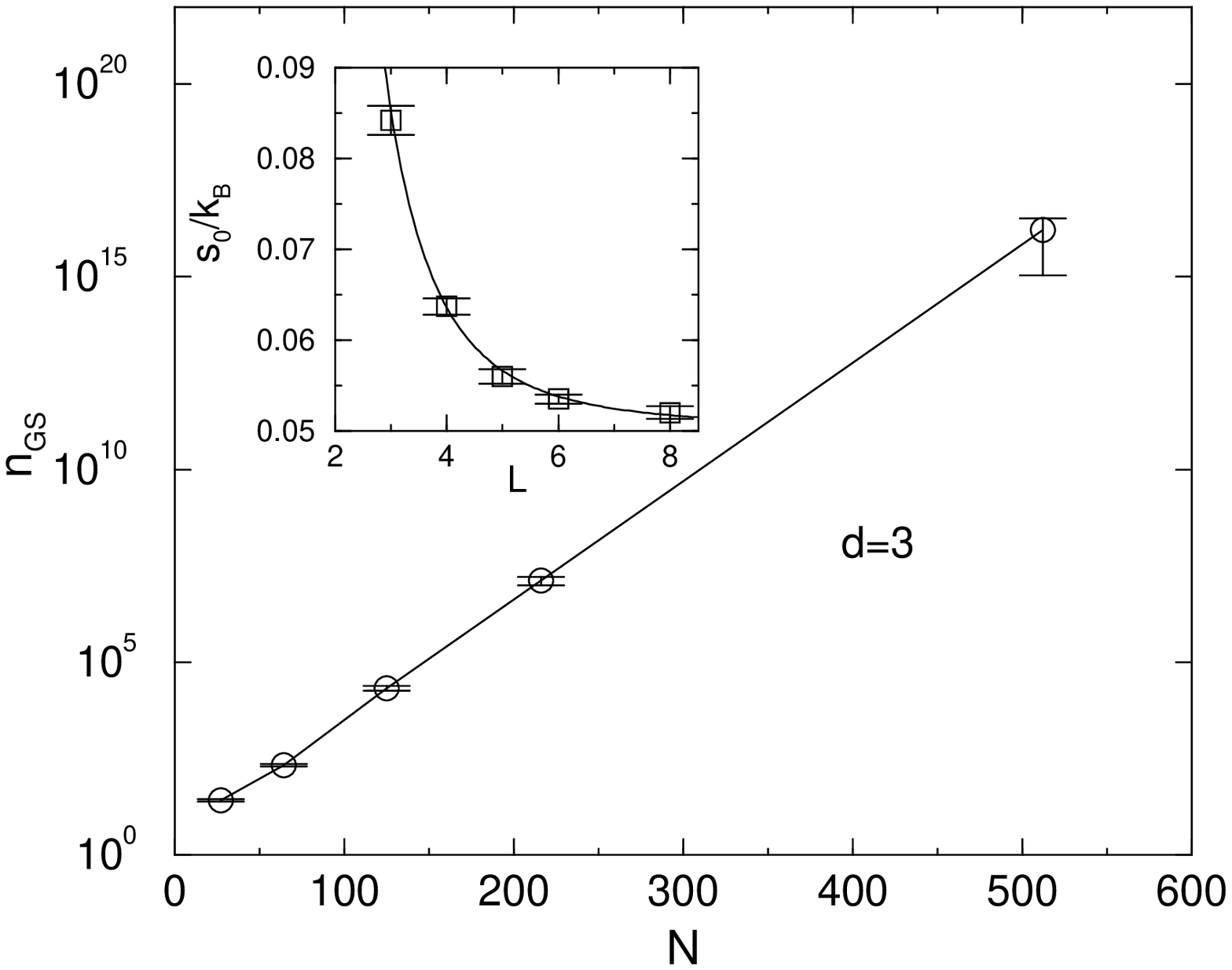}}
\end{center}
\caption{\captionNumStates}
\label{figNumStates}
\end{figure}

\begin{figure}[htb]
\begin{center}
\myscalebox{\includegraphics{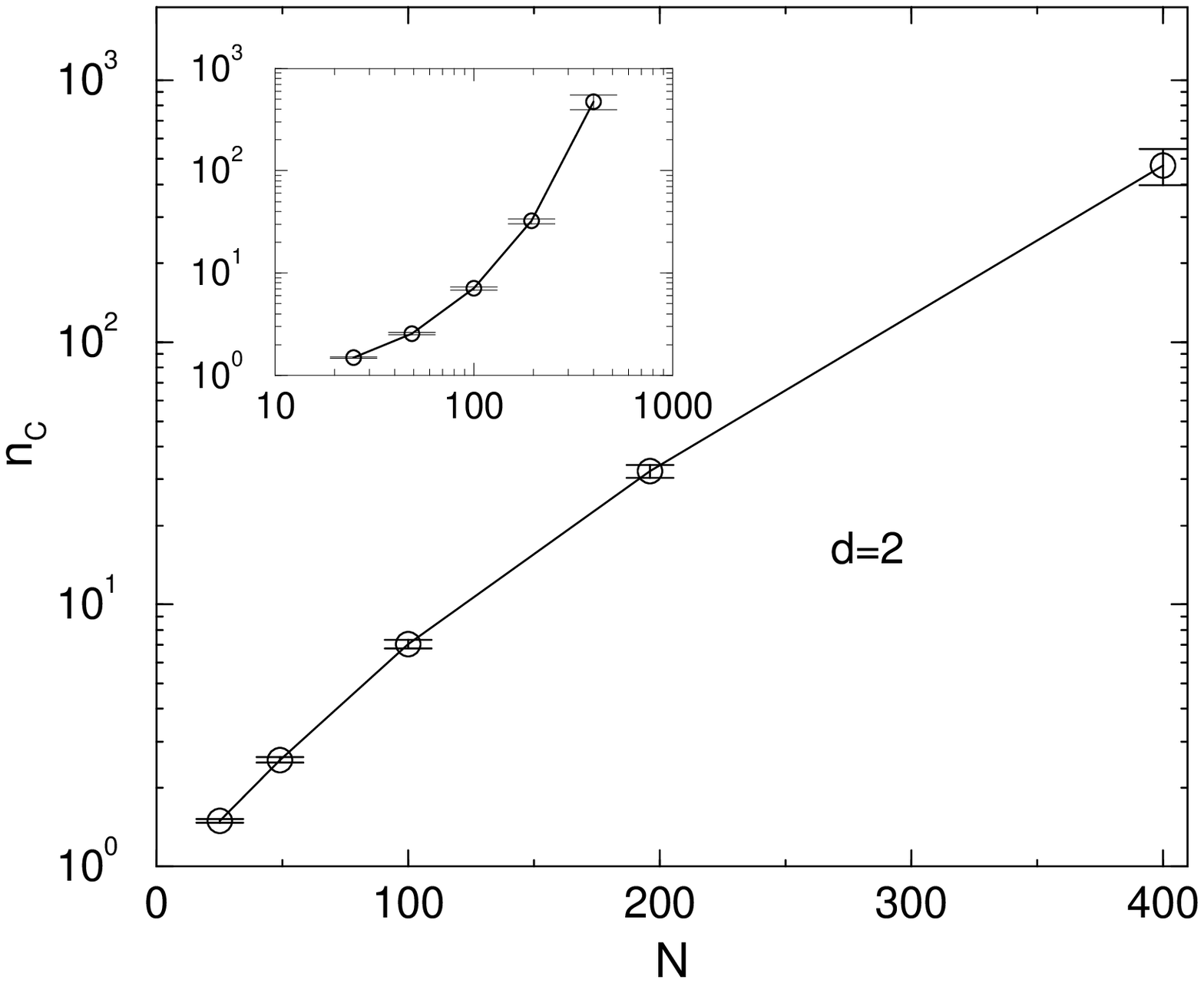}}
\end{center}
\caption{\captionNumClustersTwoD}
\label{figNumClustersTwoD}
\end{figure}

\begin{figure}[htb]
\begin{center}
\myscalebox{\includegraphics{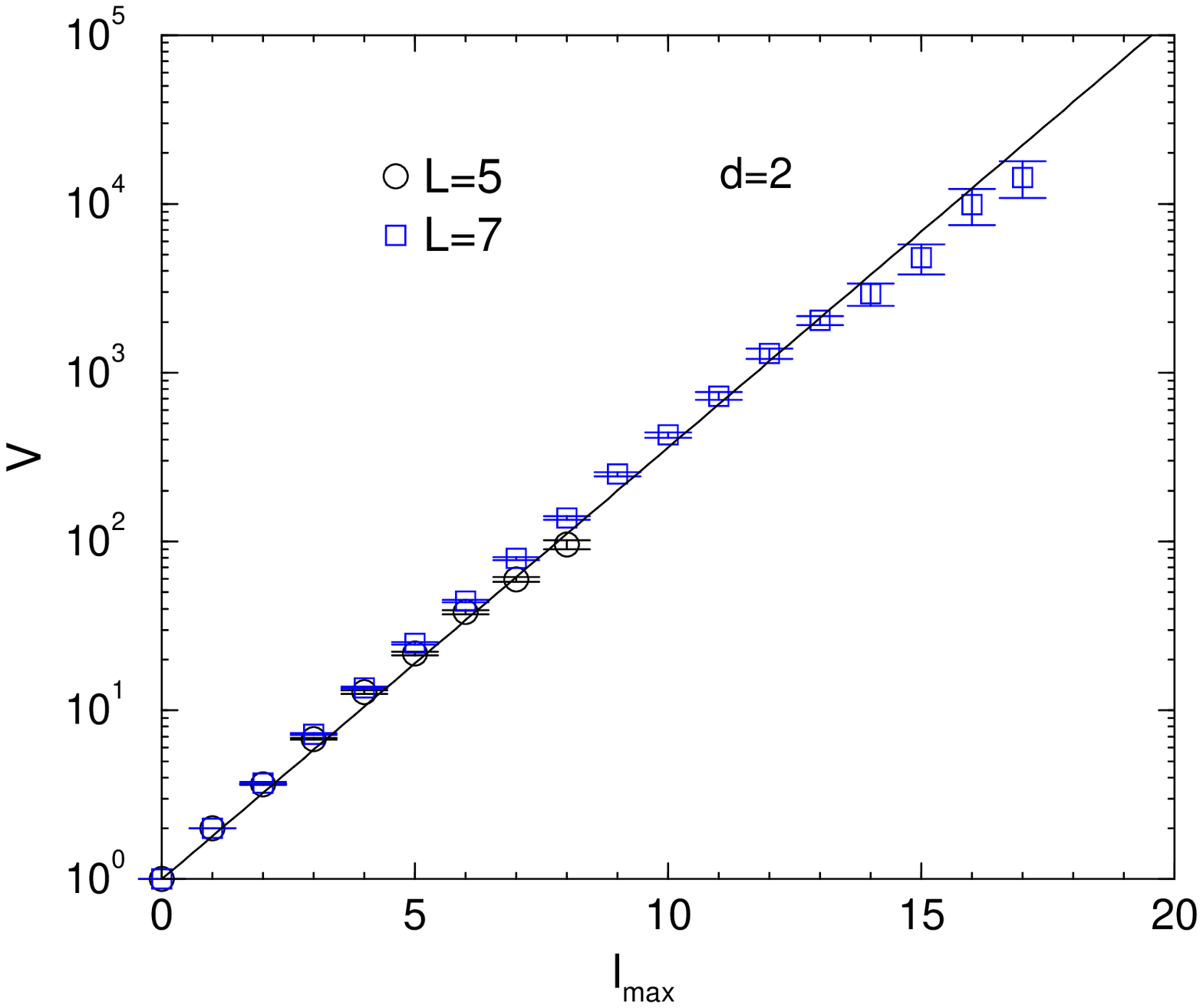}}
\end{center}
\caption{\captionEstSizeTwoD}
\label{figEstSizeTwoD}
\end{figure}

\begin{figure}[htb]
\begin{center}
\myscalebox{\includegraphics{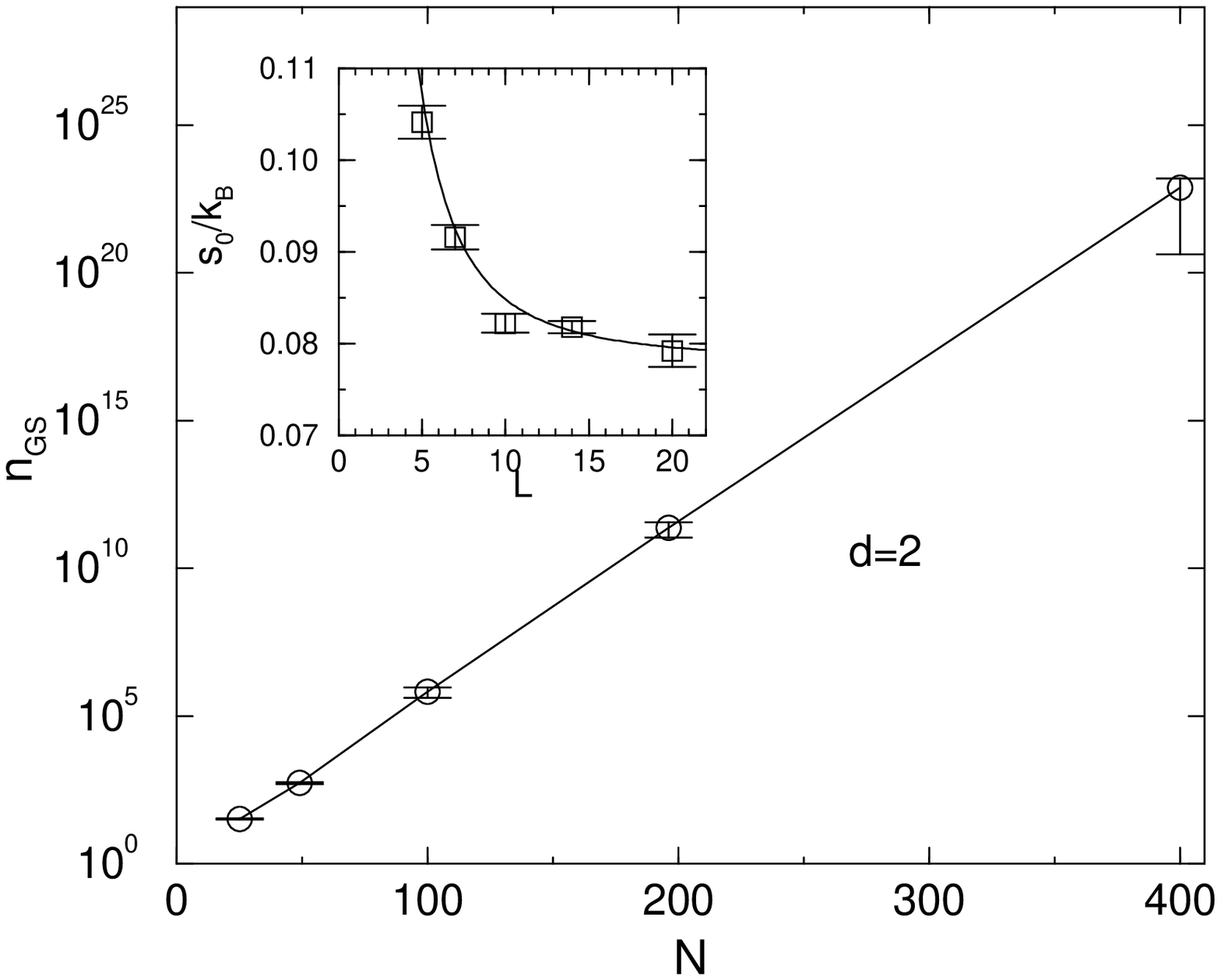}}
\end{center}
\caption{\captionNumStatesTwoD}
\label{figNumStatesTwoD}
\end{figure}

\begin{figure}[htb]
\begin{center}
\myscalebox{\includegraphics{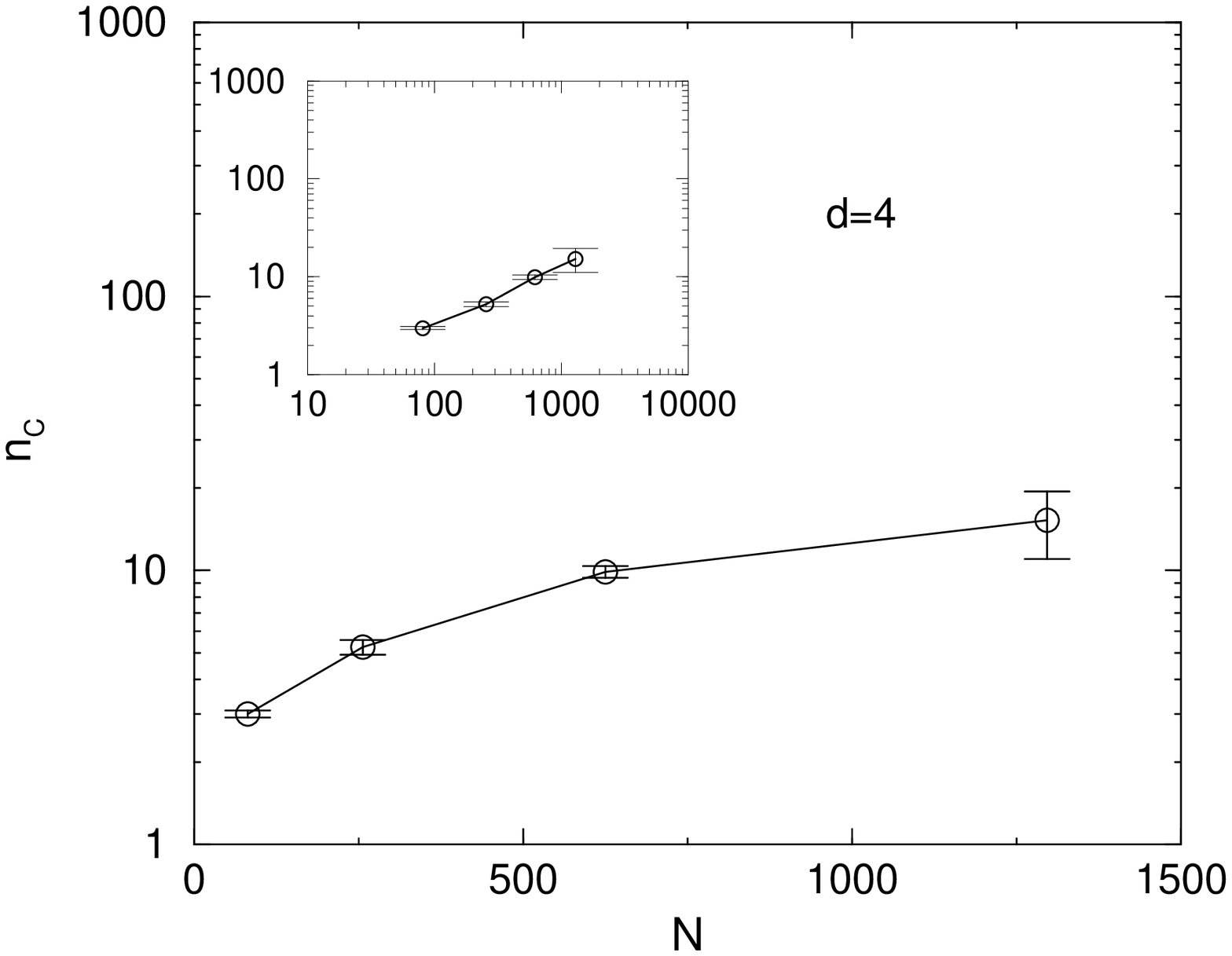}}
\end{center}
\caption{\captionNumClustersFourD}
\label{figNumClustersFourD}
\end{figure}

\begin{figure}[htb]
\begin{center}
\myscalebox{\includegraphics{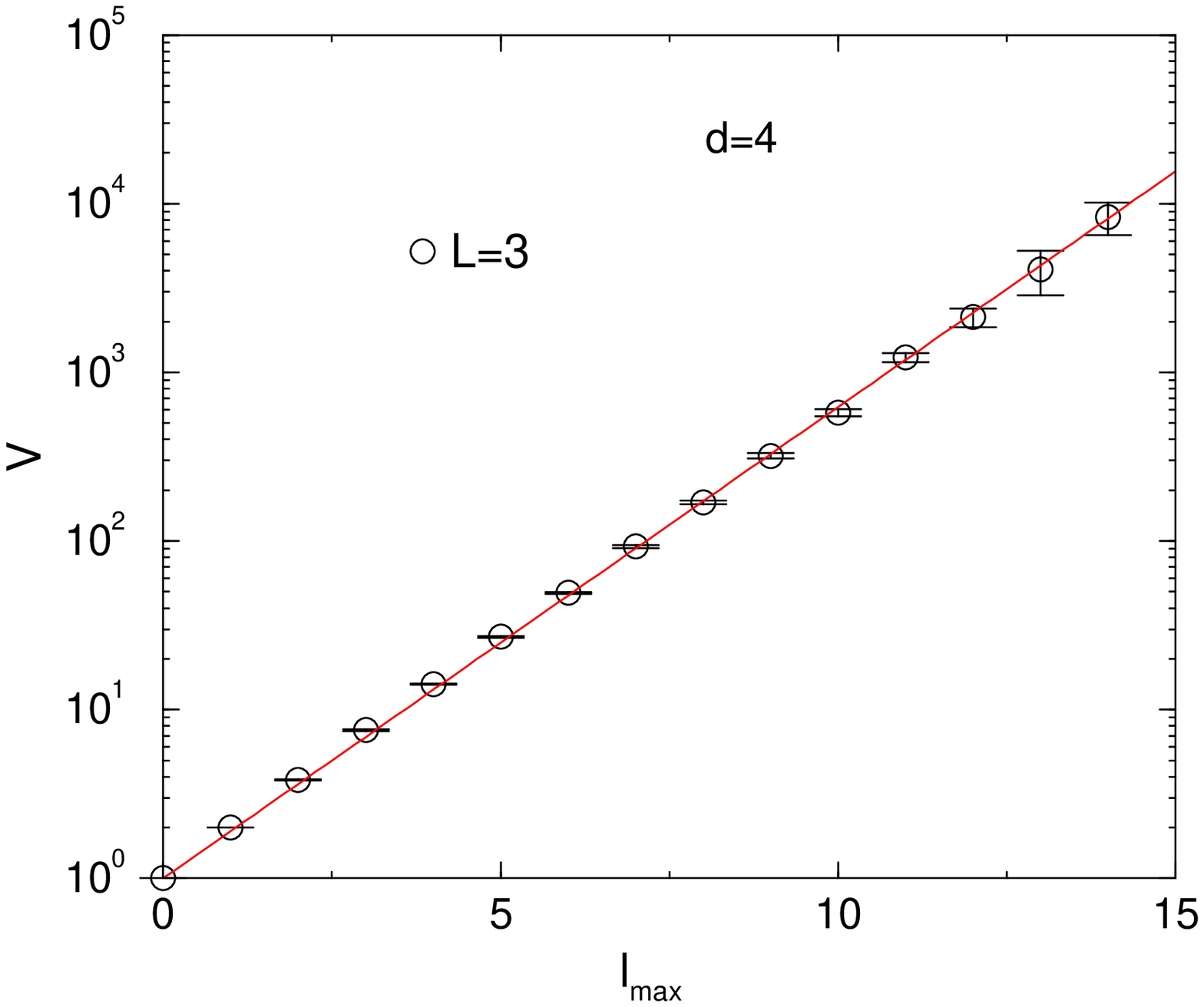}}
\end{center}
\caption{\captionEstSizeFourD}
\label{figEstSizeFourD}
\end{figure}

\begin{figure}[htb]
\begin{center}
\myscalebox{\includegraphics{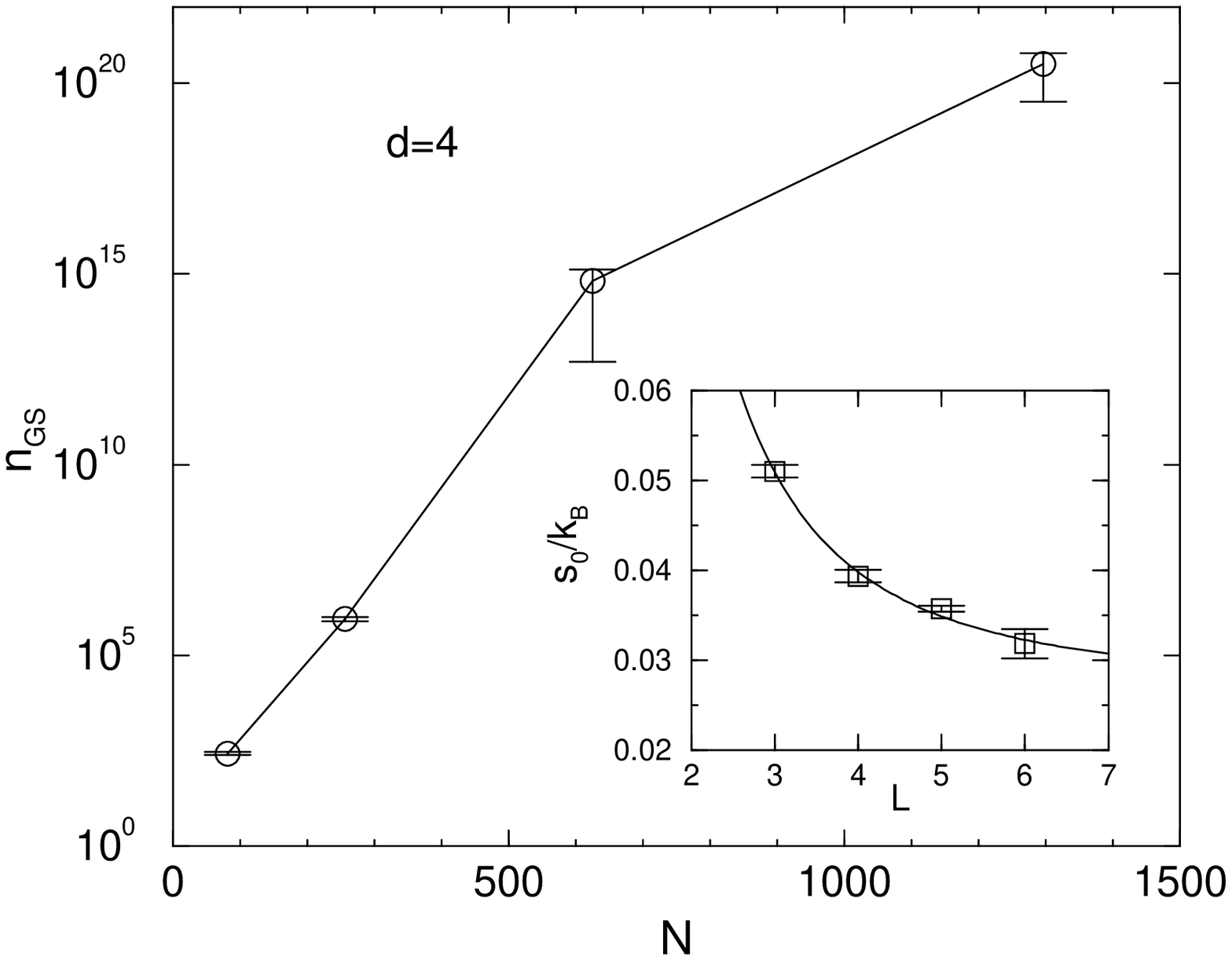}}
\end{center}
\caption{\captionNumStatesFourD}
\label{figNumStatesFourD}
\end{figure}

\end{document}